\documentclass[12pt]{article}
\usepackage{amsmath,epsfig,graphicx,amssymb}
\usepackage[margin=0.7in]{geometry}
\newcommand{\beq}{\begin{equation}}
\newcommand{\eeq}{\end{equation}}
\newcommand{\ben}{\begin{eqnarray}}
\newcommand{\een}{\end{eqnarray}}
\newcommand{\bes}{\begin{subequations}}
\newcommand{\ees}{\end{subequations}}
\newcommand{\bFig}{\begin{figure}}
\newcommand{\eFig}{\end{figure}}

\date{}
\begin{document}
\title{Continuous Transitions Between Quantum and Classical Electrodynamics}
\author{Partha Ghose\footnote{partha.ghose@gmail.com} \\
The National Academy of Sciences, India,\\ 5 Lajpatrai Road, Allahabad 211002, India}
\maketitle
\begin{abstract}
The Maxwell equations in the presence of sources are first derived without making use of the potentials and the Hamilton-Jacobi equation for classical electrodynamics is written down. The manifestly gauge invariant theory is then quantized to write down the Hamilton-Jacobi equation in quantum electrodynamics. Finally, an interpolating field theory is proposed that describes continuous transitions between quantum and classical electrodynamics. It is shown that energy flow lines are identical for quantum and classical fields in the case of the double-slit arrangement but differ in the case of vortex beams.
\end{abstract}
\section{Classical Electrodynamics}
\subsection{Maxwell Equations in Hamiltonian Form}
Consider the conventional free Hamiltonian in classical electrodynamics  
\beq
H_0[E,B] = \int d^3 x^\prime \rho_0^\prime = \int d^3 x^\prime \frac{1}{2}\left[\epsilon_0 \vec{E}^\prime.\vec{E}^\prime + \mu_0^{-1}\vec{B}^\prime.\vec{B}^\prime \right]\label{ro}
\eeq
where $E^\prime_i \equiv E_i(\vec{x}^\prime,t^\prime)$ and $B^\prime_i\equiv B_i(\vec{x}^\prime,t^\prime)$ are the electric and magnetic fields. To obtain the full set of Maxwell equations {\em in the presence of sources}, consider the change in the Hamiltonian
\ben
H[E,B,t] &=& H_0[E,B] + \int_0^t \vec{E}.\vec{j}_q d^3 x dt = \int d^3 x^\prime \rho^\prime = \int d^3 x^\prime\int_0^t \frac{\partial \rho^\prime}{\partial t^\prime}dt^\prime \label{a}\\
&=& \int d^3 x^\prime \int_0^t\left[\mu_0^{-1}\vec{B}^\prime.\frac{\partial \vec{B}^\prime}{\partial t^\prime} + \epsilon_0\vec{E}^\prime.\left(\frac{\partial \vec{E}^\prime}{\partial t^\prime} + c^2\mu_0\vec{j}_q^\prime\right)\right]dt^\prime
\een
where $\vec{E}.\vec{j}_q dt$ is the energy density absorbed by a charge $q$ in time $dt$ due to the Lorentz force acting on it.
Now define the Poynting vector $\vec{S} = \mu_0^{-1}(\vec{E}\times\vec{B})$ and the functional
\ben
\frac{\partial S[E,B,t]}{\partial t} &=& \int d^3 x^\prime \int_0^t dt^\prime\vec{\nabla}^\prime. \vec{S}^\prime \label{b}\\
&=& \mu_0^{-1}\int d^3 x^\prime \int_0^t dt^\prime\left[\vec{B}^\prime.(\vec{\nabla}^\prime\times \vec{E}^\prime) - \vec{E}^\prime.(\vec{\nabla}^\prime\times \vec{B}^\prime)\right]
\een
where use has been made of the relation $\epsilon_0\mu_0 = c^{-2}$ and the vector identity
\beq
\vec{\nabla}.(\vec{a}\times \vec{b}) = \vec{b}.(\vec{\nabla}\times\vec{a}) - \vec{a}.(\vec{\nabla}\times\vec{b}).
\eeq
Let us now define a new functional
\beq
K = H[E,B,t] + \frac{\partial S}{\partial t} = 0.\label{HJ}
\eeq
If this is to hold for all functions $(E,B)$, the Maxwell equations
\ben
\vec{\nabla} \times \vec{E} &=& - \frac{\partial \vec{B}}{\partial t},\label{m1}\\
\vec{\nabla} \times \vec{B} &=& \frac{1}{c^2}\frac{\partial \vec{E}}{\partial t} + \mu_0 \vec{j}_q,\label{m2}
\een
as well as the equation 
\beq
\frac{\partial \rho}{\partial t} + \vec{\nabla}. \vec{S} = 0,
\eeq 
or
\beq
\frac{\partial \rho_0}{\partial t} + \vec{\nabla}. \vec{S} = - \vec{E}.\vec{j}_q
\eeq 
must be satisfied. The last equation follows from eqns.(\ref{a}) and (\ref{b}) and will be recognized as Poynting's theorem. 

This completes the derivation of the full set of Maxwell equations in the presence of sources without the need to introduce the vector potential $A_\mu$. The procedure is manifestly gauge invariant. 

Now consider the free-field Maxwell equations
\ben
\vec{\nabla} \times \vec{B} &=& \frac{1}{c^2}\frac{\partial \vec{E}}{\partial t},\label{M2}\\
\vec{\nabla} \times \vec{E} &=& - \frac{\partial \vec{B}}{\partial t}\label{M1}.
\een
Notice that these equations can be written in the forms
\ben
\dot{E}_i &=& \frac{1}{\epsilon_0}\epsilon_{ijk}\partial_j \frac{\delta H_0[E,B]}{\delta B_k} = c^2\epsilon_{ijk}\partial_j B_k,\label{H1}\\
\dot{B}_i &=& -\frac{1}{\epsilon_0}\epsilon_{ijk}\partial_j \frac{\delta H_0[E,B]}{\delta E_k} = -\epsilon_{ijk}\partial_j E_k.\label{H2}
\een
These are therefore generalizations of Hamilton's equations in classical mechanics appropriate for vector field theories like Maxwell's Electrodynamics.
They can also be written in terms of Poisson brackets if the latter are generalized as follows:
\ben
\{E_i, H[E,B] \}_{PB} &=& \frac{1}{\epsilon_0}\epsilon_{ijk}\partial_j \frac{\delta H[E,B]}{\delta B_k} = \dot{E}_i,\label{H11}\\
\{B_i, H[E,B] \}_{PB} &=& -\frac{1}{\epsilon_0}\epsilon_{ijk}\partial_j \frac{\delta H[E,B]}{\delta E_k} = \dot{B}_i.\label{H22}
\een
With these generalizations $\vec{E}$ and $\vec{B}$ can be regarded as canonically conjugate variables.

\subsection{Hamilton-Jacobi Equation in Classical Electrodynamics}
For free fields $\vec{E} = \vec{E}_0 e^{i\omega t}$, eqn.(\ref{HJ}) becomes
\beq
H_0[E,B] + \frac{\partial S}{\partial t} = 0 \label{HJced}
\eeq 
with $H_0[E,B]$ given by (\ref{ro}).
Setting
\beq
E_i = \frac{1}{\omega\epsilon_0}\epsilon_{ijk}\partial_j\left(\frac{\delta S}{\delta B_k}\right) \equiv \frac{1}{\omega\epsilon_0}\left(\frac{\Delta S}{\Delta B}\right)_i,
\eeq
in equation (\ref{HJced}), one obtains
\ben
&&\frac{\partial S}{\partial t} + \frac{1}{2} \int d^3 x^\prime\left[\frac{1}{\omega^2\epsilon_0}\left(\frac{\Delta S}{\Delta B^\prime}\right)^2 + \mu_0^{-1} \vec{B}^\prime.\vec{B}^\prime\right] = 0,\label{HJcl}\\
\left(\frac{\Delta S}{\Delta B^\prime}\right)^2  &\equiv& \left[\partial^{\prime 2}\left(\frac{\delta S}{\delta B_k^\prime}\right)^2 - \partial^\prime_j\partial^\prime_k \left(\frac{\delta S}{\delta B^\prime_j}\right)\left(\frac{\delta S}{\delta B^\prime_k}\right)\right].  
\een 
This is the Hamilton-Jacobi equation in classical electrodynamics.

\section{Hamilton-Jacobi Equation in Quantum Electrodynamics}
Choosing the operators $\hat{B}_i = B_i$ and 
\beq
\hat{E}_i(\vec{x},t) = \frac{i\hbar}{\omega\epsilon_0}\epsilon_{ilm}\partial_l\frac{\delta}{\delta B_m(\vec{x},t)},\label{Eop}
\eeq
one has the equal-time commutation relations
\ben
\left[\hat{E}_i(\vec{x}, t) , \hat{B}_j(\vec{y}, t)\right] &=& -\frac{i\hbar}{\epsilon_0}\epsilon_{ijk}\partial_k \delta^3(\vec{x} - \vec{y}),\\
\left[\hat{E}_i(\vec{x}, t) , \hat{E}_j(\vec{y}, t)\right] &=& 0,\\
\left[\hat{B}_i(\vec{x}, t) , \hat{B}_j(\vec{y}, t)\right] &=& 0.
\een
This is a manifestly gauge invariant form of field quantization.

Using (\ref{Eop}), the Schr\"{o}dinger equation for a field functional $\Psi[B_i,t]$ can be written as
\ben
i\hbar \frac{\partial \Psi[B_i,t]}{\partial t} &=& H \Psi[B_i,t] =\frac{1}{2}\int d^3 x^\prime \left(- \frac{\hbar^2}{\omega^2 \epsilon_0}\frac{\Delta^2}{\Delta B^{\prime 2}} + \mu_0^{-1} \vec{B}^\prime.\vec{B}^\prime \right)\Psi[B_i,t],\label{1}\\
\frac{\Delta^2}{\Delta B^{\prime 2}} &=& \left[\partial^{\prime 2}\left(\frac{\delta}{\delta B_k^\prime}\right)\left(\frac{\delta}{\delta B_k^\prime}\right) 
- \partial^\prime_j\partial^\prime_k \left(\frac{\delta}{\delta B^\prime_j}\right)\left(\frac{\delta}{\delta B^\prime_k}\right)\right]. 
\een
Writing $\Psi$ in the polar form
\ben
\Psi[B_i,t] &=& R[B_i, t]\,{\rm exp}\left(\frac{i}{\hbar}S[B_i,t]\right) 
\een
where $R[B_i, t]$ and $S[B_i, t]$ are two real functionals, separating the real and imaginary parts, and defining
\beq
Q^\prime = - \frac{\hbar^2}{\omega^2 \epsilon_0}\frac{\Delta^2 R/\Delta B^{\prime 2}}{R},\label{4}
\eeq
one obtains the two equations
\ben
\frac{\partial S}{\partial t} + \frac{1}{2} \int d^3 x^\prime\left[\frac{1}{\omega^2\epsilon_0}\left(\frac{\Delta S}{\Delta B^\prime}\right)^2 + \mu_0^{-1} \vec{B}^\prime.\vec{B}^\prime + Q^\prime \right] &=& 0,\label{2}\\
\frac{\partial R^2}{\partial t} + \frac{1}{\omega^2\epsilon_0}\int d^3 x^\prime\frac{\Delta}{\Delta B^\prime_i} \left(R^2 \frac{\Delta S}{\Delta B^\prime_i} \right) &=& 0.\label{3}
\een
The first equation is the Hamilton-Jacobi equation in quantum electrodynamics, and the second one is a functional continuity equation for $R^2$. 

Comparing eqn.(\ref{2}) with (\ref{HJcl}), we see that it differs from (\ref{HJcl}), the classical Hamilton-Jacobi equation, only by the term $Q^\prime$. A great advantage of this is therefore that {\em all} quantum effects arising from the operator nature of the field $\hat{E}_i$ are neatly isolated in the quantum potential $Q$ alone.  
 
Defining the momentum $\Pi_i =\frac{1}{\omega^2\epsilon_0}\Delta S/\Delta B_i \equiv E_i$, one can rewrite eqn. (\ref{2}) in the form
\ben
&&\frac{\partial S}{\partial t} + \frac{1}{2} \int d^3 x^\prime\left[\epsilon_0 \vec{\Pi}.\vec{\Pi} + \mu_0^{-1}\vec{B}.\vec{B} + Q^\prime\right]\nonumber\\
&\equiv&\frac{\partial S}{\partial t} + \frac{1}{2} \int d^3 x^\prime\left[\epsilon_0 \vec{E}.\vec{E} + \mu_0^{-1}\vec{B}.\vec{B} + Q^\prime\right]\nonumber\\
&=& \frac{\partial S}{\partial t} + \int d^3 x^\prime\rho_{Q}^\prime = \frac{\partial S}{\partial t} + \int d^3 x^\prime \left(\rho_{cl}^\prime + Q^\prime\right) =  0.
\een
Following the standard procedure, one can therefore show that the Poynting vector $\vec{S}$ satisfies the equation
\ben
\frac{\partial \rho_Q}{\partial t} + \vec{\nabla}. \vec{S} = 0, 
\een
or
\ben
\frac{\partial {\rho}_{cl}}{\partial t} + \vec{\nabla}. \vec{S} = - \frac{\partial Q}{\partial t},\label{poy} 
\een
which is the Poynting theorem in quantum electrodynamics. 
\section{An Interpolating Field}
Let us now postulate an interpolating nonlinear Schr\"{o}dinger equation 
\beq
i\hbar \frac{\partial \Psi[B_i,t]}{\partial t} = \frac{1}{2}\int d^3 x^\prime\left(- \frac{\hbar^2}{\omega^2 \epsilon_0}\frac{\Delta^2}{\Delta B^{\prime2}} + \mu_0^{-1} \vec{B}^\prime.\vec{B}^\prime -\lambda(\theta)Q^\prime \right)\Psi[B_i,t]\label{5}
\eeq
where $\lambda(\theta)$ is a coupling function that is assumed to be a continuous function of some parameter $\theta$, satisfying the conditions $0 < \lambda(\theta) \leq 1$. Notice that $\lambda(\theta)\rightarrow 0$ is the quantum limit (eqn. (\ref{2})) and $\lambda(\theta)\rightarrow 1$ is the classical limit in which the quantum potential density $Q$ does not contribute. Then the modified functional Hamilton-Jacobi equation for a `mesostate' of the electromagnetic field is
\ben
\frac{\partial S}{\partial t} + \int d^3 x^\prime \rho^\prime_{mes}  &=& 0,\label{6}
\een
with
\ben
\rho_{mes} &=& \frac{1}{2}\left[\epsilon_0 \vec{E}.\vec{E} + \mu_0^{-1}\vec{B}.\vec{B} + (1 - \lambda(\theta)) Q\right] \label{rhomes}\\
&=& \rho_{cl} + (1 - \lambda(\theta)) Q.
\een
Thus, equation (\ref{5}) provides a basis for studying continuous transitions between quantum and classical electrodynamics. 

The Poynting theorem now takes the form
\ben
\frac{\partial \rho_{mes}}{\partial t} + \vec{\nabla}. \vec{S} = 0. 
\een 
Using the relation
\beq
\vec{S} = \rho_{mes}\vec{c},
\eeq
we get
\beq
\vec{c}= c\frac{d \vec{r}}{ds} = \frac{\vec{S}}{\rho_{mes}} \label{diff}
\eeq
where $ds$ is the infinitesimal arc length corresponding to an infinitesimal path $d\vec{r}$ of energy flow. This differential equation determines the energy-flow pattern in the field, and as is evident, this pattern changes with the parameter $\lambda(\theta)$, giving the classical pattern in the limit $\lambda(\theta)\rightarrow 1$ and the fully quantized pattern in the limit $\lambda(\theta)\rightarrow 0$. 

This is thus a generalization of the theory of continuous transitions between quantum and classical mechanics formulated earlier \cite{ghose1, ghose2, ghose3} to the case of a relativistic field theory. 

\section{Applications}
\subsection{The Double-slit}
The energy flow lines for plane ploarized light behind a two-dimensional Ronchi grating have been calculated in Ref. \cite{dav}. Consider a monochromatic solution of the Maxwell equations with frquency $\omega$ given by
\ben
\vec{E}(\vec{r}, t) &=& \vec{E}(\vec{r}) e^{-i\omega t},\\
\vec{H}(\vec{r}, t) &=& \vec{H}(\vec{r}) e^{-i\omega t}.   
\een
It is well known that the time independent parts of the fields satisfy the Helmholtz equations
\ben
\nabla^2 \vec{E}(\vec{r}) + k^2\vec{E}(\vec{r}) &=& 0,\\
\nabla^2 \vec{H}(\vec{r}) + k^2\vec{H}(\vec{r}) &=& 0, 
\een
where $k = \omega/c$.

Now consider a linearly polarized plane light wave propagating in the $y$ direction and incident on a grating in the $xz$ plane at $y=0$. Then, as shown in Born and Wolf \cite{bw} and also Ref. \cite{dav}, the fields diffracted by the grating are given by
\ben
\vec{H} &=& \mp ik^{-1} Be\left(\frac{\partial \psi}{\partial y}\hat{e}_x -\frac{\partial \psi}{\partial x}\hat{e}_y \right) + A\psi\hat{e}_z,\\
\vec{E} &=& \frac{iA}{\epsilon_0 \omega}\left(\frac{\partial \psi}{\partial y}\hat{e}_x - \frac{\partial \psi}{\partial x}\hat{e}_y \right) \pm \frac{kB}{\epsilon_0 \omega}\psi\hat{e}_z,\label{E}
\een
where $A$ and $B$ are real constants and $\psi$ is a solution of the Helmholtz equation
\beq
(\nabla^2 + k^2)\psi(x,y) =0,
\eeq
satisfying the grating boundary conditition. The components of the time averaged Poynting vector are given by
\ben
S_x &=& \frac{i}{4\epsilon_0\omega}(A^2 + B^2)\left(\psi \frac{\partial \psi^*}{\partial x} -\psi^* \frac{\partial \psi}{\partial x}\right),\\
S_y &=& \frac{i}{4\epsilon_0\omega}(A^2 + B^2)\left(\psi \frac{\partial \psi^*}{\partial y} -\psi^* \frac{\partial \psi}{\partial y}\right),\\
S_z &=& 0.
\een
Hence, the energy flow pattern determined by eqn. (\ref{diff}) is given by 
\beq
\frac{dy}{dx} = \frac{S_y}{S_x} = \frac{\left(\psi \frac{\partial \psi^*}{\partial y} -\psi^* \frac{\partial \psi}{\partial y} \right)}{\left(\psi \frac{\partial \psi^*}{\partial x} -\psi^* \frac{\partial \psi}{\partial x}\right)}.
\eeq
Since $\rho_{mes}$ drops out of this ratio, {\em the energy flow pattern is the same for all values of $\lambda(\theta)$}. This is an important result of relevance for the interpretation of empirically determined energy flow patterns \cite{stein}. 

\subsection{Optical Vortices}
The easy production of light beams with discrete angular momentum in the laboratory using laser beams has led to novel experiments to manipulate microparticles \cite{padget}. Lasers typically produce Hermite-Gauss beams. Helical Laguerre-Gauss beams can be readily produced by conversion of these Hermite-Gauss modes using specially designed holograms. They have helical phase fronts because the Poynting vector is not parallel to the beam axis but follows a spiral trajectory, creating `optical vortices' with zero intensity along the beam axis. 

The paraxial approximation of the Helmholtz solutions is the most appropriate to desctibe these modes. For linearly polarized light the paraxial fields are given by \cite{padget1}
\ben
\vec{B} &=& = ik\left[u\hat{y} + \frac{i}{k}\frac{\partial u}{\partial y}\hat{z}\right] e^{ikz},\label{B2}\\
\vec{E} &=& ik\left[u\hat{x} + \frac{i}{k}\frac{\partial u}{\partial x}\hat{z}\right]e^{ikz},\label{E2}
\een
where the propagation is now taken along the $z$ axis, $u$ being the complex scalar function that describes the distribution of the field amplitude in the paraxial approximation. In the Lorentz gauge and in cylindrical coordinates, this function is given by \cite{padget2}
\ben
u_{pl}(r,\phi,z) &=& \frac{C}{(1+z^2/z^2_R)^{1/2}}\left[\frac{r\sqrt{2}}{w(z)}\right]^l L_p^l\left[\frac{2r^2}{w^2(z)} \right]\nonumber\\&\times&{\rm exp}\left[\frac{-r^2}{w^2(z)} \right]
{\rm exp}\left(\frac{-ikr^2z}{2(z^2 + z_R^2)}\right){\rm exp}\left(-il\phi\right)\nonumber\\&\times& {\rm exp}\left[i(2p + l +1){\rm tan}^{-1}\frac{z}{z_R} \right],\label{LG}
\een
where $C$ is a constant, $z_R$ is the Rayleigh range, $w(z)$ is the radius of the beam, $L_p^l$ is the associated Laguerre polynomial, and the beam waist is at $z=0$. The time averaged Poynting vector for a beam of unit amplitude is then given by
\ben
\vec{S} =\mu_0^{-1} (\vec{E} \times \vec{B}) &=& \frac{1}{2\mu_0}\left[\langle\vec{E}^* \times \vec{B}\rangle + \langle\vec{E} \times \vec{B}^*\rangle\right]\nonumber\\
&=& i\omega \frac{1}{2\mu_0}\left(u\nabla u^* - u^*\nabla u\right) + \frac{\omega}{\mu_0} k|u|^2 \hat{z}.
\een
Using the LG mode given by (\ref{LG}) for linearly polarized beams, one can show that \cite{padget2}
\ben
\vec{S} = c\left[\frac{rz}{z^2 + z_R^2}|u|^2\hat{r} + \frac{l}{kr}|u|^2\hat{\phi} + |u|^2\hat{z}\right]
\een
where $|u|^2 = |u(r,\phi,z)|^2$ and the hats denote unit vectors. This shows that the Poynting vector spirals along the direction of propagation $\hat{z}$. 

One can calculate $\rho_{mes}$ from (\ref{rhomes}), (\ref{B2}), (\ref{E2}) and the quantum potential is given by
\ben
Q &=& - \frac{\hbar^2}{\omega^2 \epsilon_0}\frac{\Delta^2 |u(r,\phi,z)|/\Delta B^2}{|u(r,\phi,z|}.
\een
One can then compute the energy flow lines from eqn. (\ref{diff}) and study the continuous transtions between the quantum and classical flows.
\section{Acknowledgement}
The author thanks the National Academy of Sciences, India for a grant. He also specially thanks Anirban Mukherjee for a very helpful discussion and Riddhi Chatterjee for carefully checking the calculations.

\end{document}